\documentclass[conference]{IEEEtran}

\usepackage{amsfonts, amsmath, amsthm, amssymb, bm}
\usepackage{mathtools}
\usepackage{graphicx}
\usepackage{tikz, standalone}

\usepackage{algorithm}
\usepackage[noend]{algpseudocode}
\usepackage[caption=false,font=footnotesize]{subfig}

\usepackage{booktabs}
\usepackage{physics}
\usepackage{siunitx}

\usepackage[table]{xcolor}
\definecolor{rowgray}{gray}{0.92} 

\IEEEoverridecommandlockouts
\allowdisplaybreaks
\begin{document}
\title{
Dual-Diode Unified SWIPT for High Data Rates with Adaptive Detection\thanks{This work has received funding from the European Union’s Horizon Europe programme (ERC, WAVE, Grant agreement No. 101112697).}
}

\author{
\IEEEauthorblockN{Zulqarnain Bin Ashraf\IEEEauthorrefmark{1}, Triantafyllos Mavrovoltsos\IEEEauthorrefmark{2},  Constantinos Psomas\IEEEauthorrefmark{3}, Ioannis Krikidis\IEEEauthorrefmark{2}, and Besma Smida\IEEEauthorrefmark{1}}
\IEEEauthorblockA{\IEEEauthorrefmark{1}Department of Electrical and Computer Engineering, University of Illinois Chicago, USA}
\IEEEauthorblockA{\IEEEauthorrefmark{2}Department of Electrical and Computer Engineering, University of Cyprus, Cyprus}
\IEEEauthorblockA{\IEEEauthorrefmark{3}Department of Computer Science and Engineering, European University Cyprus, Cyprus}
\IEEEauthorblockA{Emails: zbinas2@uic.edu, tmavro03@ucy.ac.cy,  c.psomas@euc.ac.cy, krikidis@ucy.ac.cy, smida@uic.edu}\vspace{-10mm}
}

\maketitle

\begin{abstract} Due to their low-complexity and energy-efficiency, unified simultaneous wireless information and power transfer (U-SWIPT) receivers are especially suitable for low-power Internet of Things (IoT) applications. Towards accurately modeling practical operating conditions, in this study, we provide a unified transient framework for a dual-diode U-SWIPT that jointly accounts for diode nonlinearity and capacitor-induced memory effects. The proposed model accurately describes the inherent time dependence of the rectifier, highlighting its fundamental impact on both energy harvesting (EH) and information decoding (ID) processes. Based on the provided memory-aware model, we design a low-complexity adaptive detector that learns the nonlinear state transition dynamics and performs decision-directed detection with linear complexity. The proposed detection scheme approaches maximum likelihood sequence detection (MLSD) performance in memory-dominated regimes, while avoiding the exponential search required by classical sequence detection. Overall, these results demonstrate that properly exploiting rectifier memory provides a better tradeoff between data rate and reliability for U-SWIPT receivers.
	\end{abstract}
	
	
	\section{Introduction}
	Simultaneous wireless information and power transfer
	(SWIPT) uses the same radio-frequency (RF) signal to deliver
	energy and information to low-power devices, making it a
	promising technology for powering massive IoT deployments
	in future 6G networks \cite{8476597}. Although early SWIPT implementations relied on separated receiver architectures, the unified
	(U-SWIPT) receivers have gained significant attention due to their reduced hardware complexity and improved energy efficiency \cite{10980390}. By employing a rectifying front-end in the U-SWIPT architecture, both energy harvesting (EH) and information decoding (ID) tasks are merged into a single circuit, reducing complexity and power
	consumption. However, this integration introduces strong nonlinear coupling between the EH and ID processes, leading to challenging design trade-offs \cite{10980390}. 
	
	To analyze the performance demonstrated by U-SWIPT receivers, early studies relied on simplified linear power conversion models
	that neglected diode nonlinearities \cite{6623062,6678102}. More recent work incorporated the nonlinear current-voltage characteristics of
	diodes, providing a more accurate representation of rectifier behavior \cite{7547357,7264986}. Despite these improvements, most existing analyses still treat the rectifier as a memoryless circuit, overlooking the capacitor charge and discharge dynamics that introduce memory. To this end, a few recent works explicitly account for rectifier memory. For instance, the authors in \cite{9241856} model the capacitor-induced memory using a Markov
	decision process. In addition, assuming a biased
	amplitude shift keying (BASK) modulation scheme, the subsequent works in \cite{11015328} and \cite{10622830} developed simple yet accurate memory-aware models, showing that capacitor memory causes inter-symbol interference (ISI). Their results highlight that the memory-induced distortion significantly affects ID and EH performance, revealing that maximum likelihood sequence detection (MLSD) significantly improves bit error rate (BER) performance by utilizing circuit memory. However, this improvement in the tradeoff between data rate and decoding reliability comes at the cost of high computational complexity.
	
	Conventional single-diode U-SWIPT designs exploit a single rectified path to extract energy and information, demonstrating limited performance across both EH and ID processes. In contrast, a double half-wave rectifier introduced in \cite{8664629} employs two anti-parallel diodes to utilize both positive and negative waveform polarities, further pushing the performance limits. However, their analysis assumes a steady-state regime with large capacitors and does not account for the transient memory effects that arise at shorter symbol durations. \\
    {\bf{Contribution:}} Towards a more practical and accurate scenario, this work:  
    \begin{itemize}
        \item  Develops a unified transient framework for a double half-wave rectifier that jointly captures diode nonlinearity and capacitor-induced memory in a dual-diode configuration. 
        \item Derives  a complete time-domain model  to describe all conduction states and transitions, accurately	representing the circuit dynamics across symbols. 
        \item Introduces a new circuit-aware adaptive detector, built on this model, that learns the nonlinear state evolution and performs recursive detection, achieving near-MLSD performance with only linear computational complexity.

  \end{itemize}

\section{System Model \& Circuit Analysis}
\label{sec:system_model}
\begin{figure}[!t]
	\centering
	\includegraphics[scale=0.38]{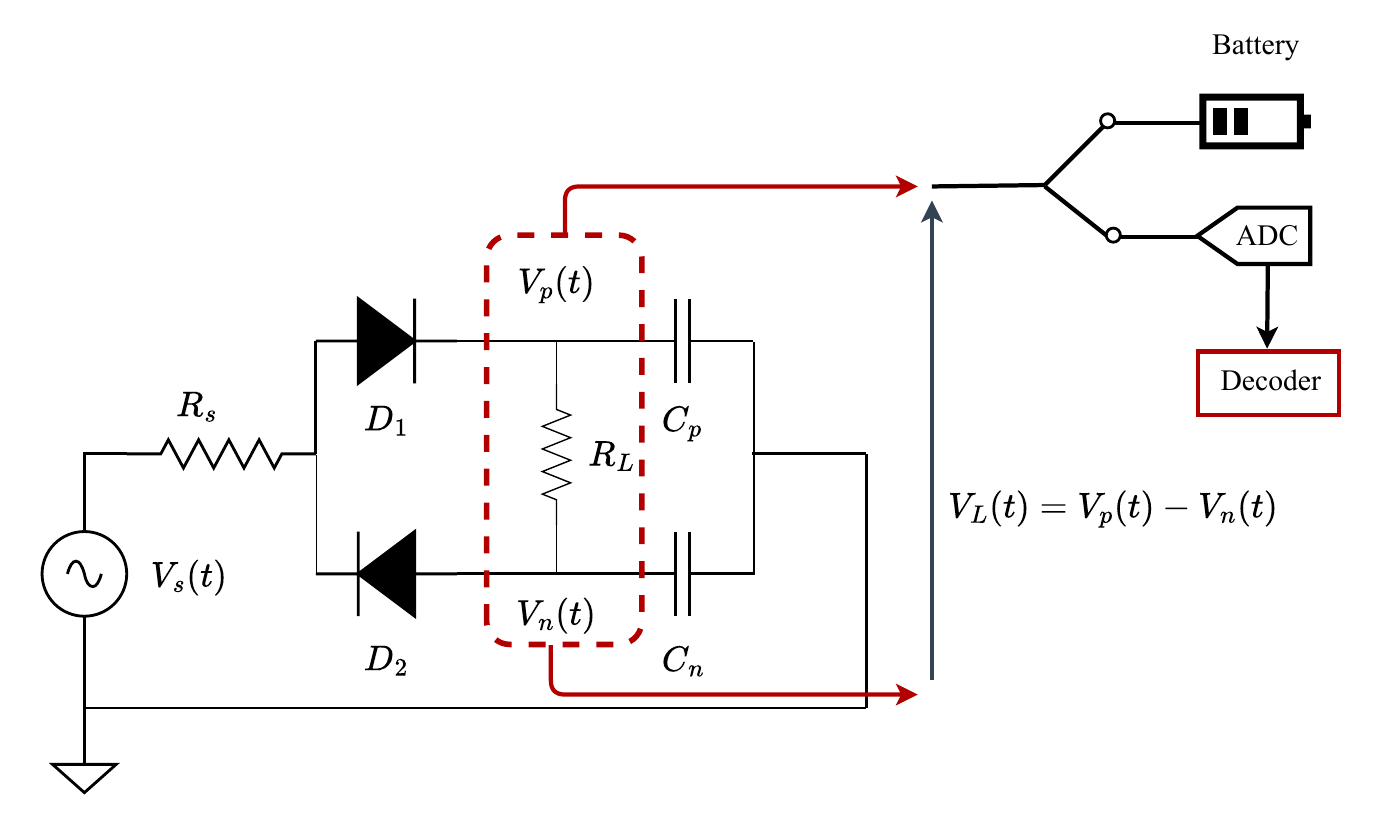}\vspace{-3mm}
	\caption{Considered U-SWIPT receiver, illustrating the nodes for EH and ID.
     Depending on the system needs and priorities, the rectifier output is selectively used for information decoding (ID) or energy harvesting (EH).}
     \vspace{-2mm}
	\label{fig:dualcktschematics}
\end{figure}

We consider a point-to-point system where the receiver jointly harvests energy and decodes information from the same RF signal. The receiver consists of two anti-parallel diodes $D_1$ and $D_2$, a load resistor $R_L$, and two low-pass filter capacitors $C_p$ and $C_n$, forming a double half-wave rectifier, as illustrated in Fig.~\ref{fig:dualcktschematics}. The diodes rectify the received signal in opposite directions, producing positive and negative rectified output voltages $V_p(t)$ and $V_n(t)$, respectively. These signals are jointly used for both EH and ID~\cite{8664629}.
	
At the transmitter, a binary BASK waveform is employed to enable both information transfer and continuous power delivery. The amplitude set $ A_k \in \{A_L, A_H\}, \quad A_H > A_L \ge A_{\min},$ is selected to maintain nonzero harvested energy and maximize the output voltage separation for reliable detection. Subsequently, the received signal during the $k$-th symbol period is modeled as
	\begin{equation}
		V_k(t) = A_{k} \sin(2 \pi f_c t + \phi),
		\label{eq:Vs}
	\end{equation}
	where $A_k$, $f_c$, and $\phi$ denote the received amplitude, carrier frequency, and phase, respectively. Without loss of generality, we set $\phi = 0$, since the rectifier discards phase information.

\subsection{Piecewise Diode Model}

To characterize the rectifier dynamics, we adopt a piecewise-linear diode model. Each diode $ D_\ell, \, \ell \in \{1,2\}$ is modeled with a turn-on voltage \(V_{\mathrm{on}} >0\), forward resistance \(R_{\mathrm{on}}\) and a reverse resistance \(R_{\mathrm{off}}\), where $R_{off}\gg R_{on}$. Each diode \(D_\ell\) is represented by a piecewise-linear current–voltage characteristic, given by
	\begin{equation}
		I_{D\ell}(t)=
		\begin{cases}
			\dfrac{V_{D\ell}(t) - V_{\mathrm{on}}}{R_{\mathrm{on}}}, & V_{D\ell}(t) \ge V_{\mathrm{on}},\\[0.9ex]
			\dfrac{V_{D\ell}(t)}{R_{\mathrm{off}}}, & V_{D\ell}(t) < V_{\mathrm{on}},
		\end{cases}
		\qquad \ell \in \{1,2\}.
		\label{eq:PL-diode-both}
	\end{equation}
	At any time, the dual-diode rectifier operates in one of four instantaneous modes
	\(\sigma\in\{\mathrm{RR},\mathrm{FR},\mathrm{RF},\mathrm{FF}\}\),
	corresponding to the bias states of  \(D_\ell\). The guard conditions for these modes, along with their effective diode resistances, are summarized in Table \ref{tab:modes_full}. In mode~\(\sigma\), the effective diode resistances are denoted by \((R_{1,\sigma}, R_{2,\sigma}) \in \{R_{\mathrm{on}}, R_{\mathrm{off}}\}\).
    
\subsection{Ordinary Differential Equations for Circuit Analysis}

 \begin{figure*}[t]
	\centering
	\scalebox{0.74}{%
		\begin{minipage}{\textwidth}
			\subfloat[FR: $D_1$ On, $D_2$ Off\label{fig:state_fr}]{
				\includegraphics[width=0.48\linewidth]{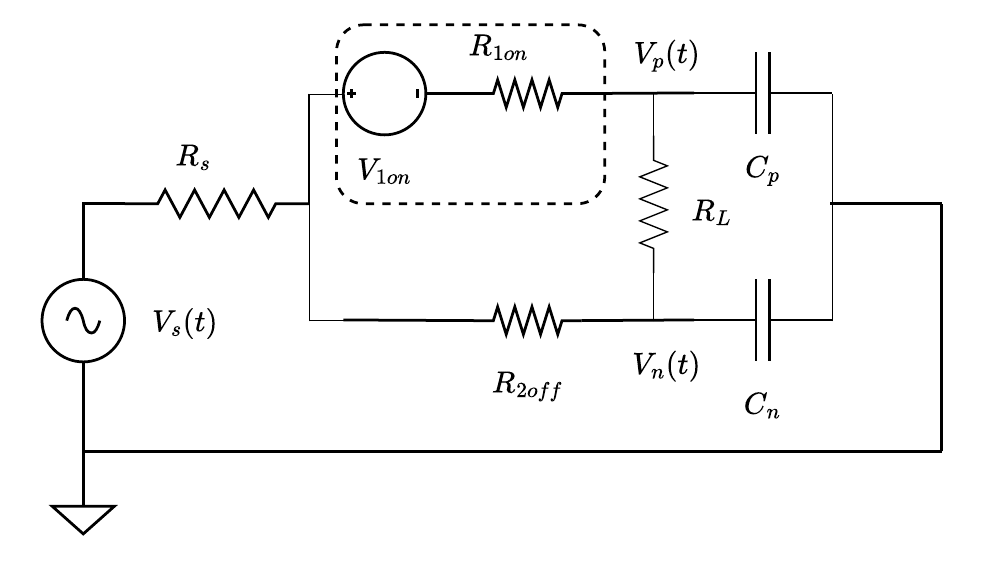}
			}\hfill
			\subfloat[RF: $D_1$ Off, $D_2$ On\label{fig:state_rf}]{
				\includegraphics[width=0.48\linewidth]{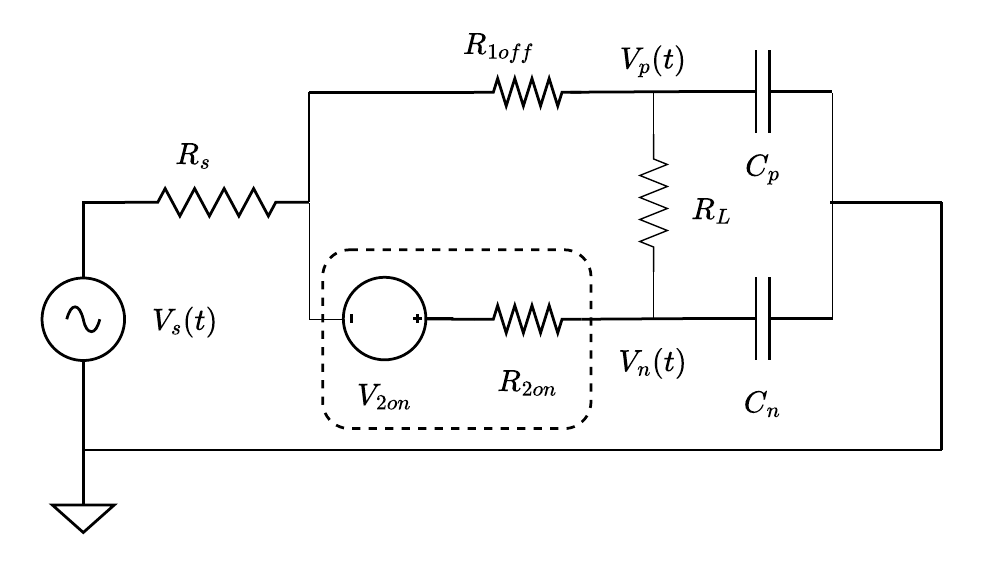}
			}
			
			\vspace{-2mm}
			
			\subfloat[RR: both Off\label{fig:state_rr}]{
				\includegraphics[width=0.48\linewidth]{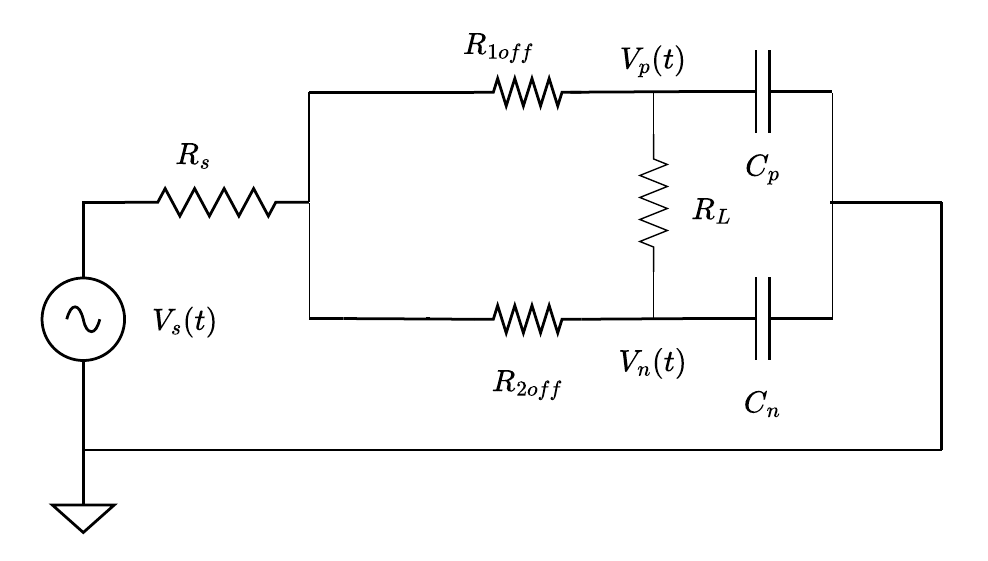}
			}\hfill
			\subfloat[FF: both On\label{fig:state_ff}]{
				\includegraphics[width=0.48\linewidth]{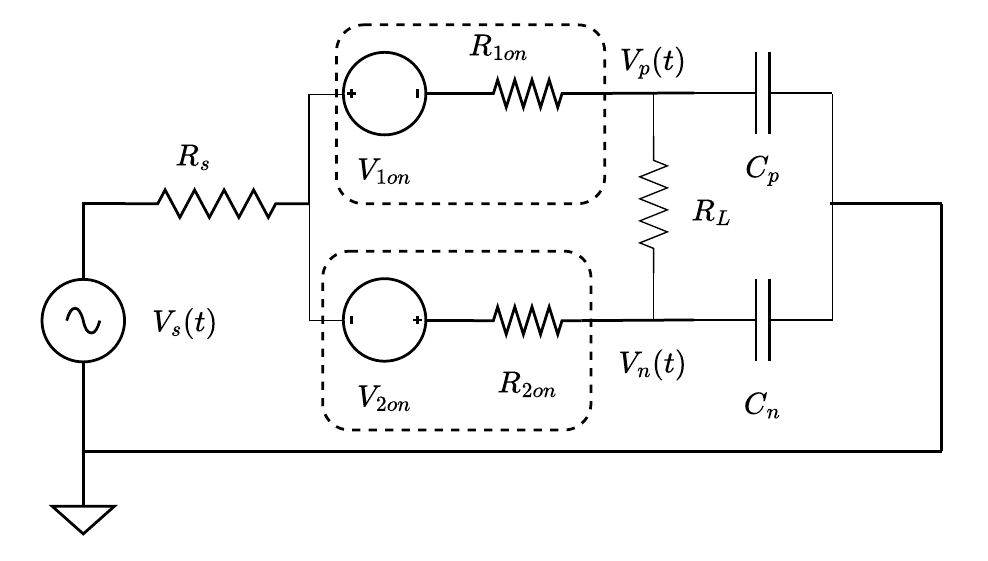}
			}
		\end{minipage}
	}
	\caption{Four conduction states of the dual-diode U-SWIPT receiver.}\vspace{-2mm}
	\label{fig:four_states_2x2}
\end{figure*}

Given the source resistance \(R_s\), the load \(R_L\), and the filter capacitors \(C_p, C_n\), Kirchhoff’s current law (KCL) at the rectified nodes yields the following coupled first-order ordinary differential equations (ODEs), expressed as
\begin{align}
(R_s+R_{1,\sigma})C_p\,\dot V_p(t)+ R_s C_n\,\dot V_n(t)+ \Big(1+\tfrac{R_{1,\sigma}}{R_L}\Big)V_p(t)&\notag\\\quad 
-\,\tfrac{R_{1,\sigma}}{R_L}V_n(t)= u_{1,\sigma}(t), \label{eq:coupled1}\\[2pt]
R_{1,\sigma}C_p\,\dot V_p(t)- R_{2,\sigma}C_n\,\dot V_n(t)+ \tfrac{R_{1,\sigma}+R_L+R_{2,\sigma}}{R_L}V_p(t)&\notag\\\quad -\,\tfrac{R_{1,\sigma}+R_L+R_{2,\sigma}}{R_L}V_n(t)= u_{2,\sigma}(t). \label{eq:coupled2}\end{align}

The resulting closed-form solution in each mode is:

\begin{align}V_p(t)= &\, C_1 e^{r_1^{(\sigma)}(t - t_0)}   + C_2 e^{r_2^{(\sigma)}(t - t_0)} \notag\\  &+ \bar V_s \!\left[a^{(\sigma)} \cos(2\pi f t)  + b^{(\sigma)} \sin(2\pi f t)\right]  + d^{(\sigma)},\label{eq:Vp_solution}\end{align} 
where $r_{1,2}^{(\sigma)}$ are the characteristic roots and$(a^{(\sigma)}, b^{(\sigma)}, d^{(\sigma)})$ denote the harmonic gains and DC offset for mode~$\sigma$. The constants $C_1$ and $C_2$ enforce continuity of $V_p(t)$ and $\dot V_p(t)$ at the mode-transition instant $t_0$. The complementary voltage $V_n(t)$ follows from back-substitution:
\begin{equation}V_n(t)= h_{1,\sigma}V_p(t)+ h_{2,\sigma}\dot V_p(t)+ h_{3,\sigma}V_s(t)+ h_{4,\sigma},\label{eq:Vn_backsub}
\end{equation}
where the coefficients $h_{i,\sigma}$ depend on $\{R_s,R_L,C_p,C_n,R_{1,\sigma},R_{2,\sigma}\}$ and the diode thresholds. The detailed algebraic derivation of Eqs.~\eqref{eq:Vp_solution}–\eqref{eq:Vn_backsub}, including the expressions for $r_{1,2}^{(\sigma)}$, $a^{(\sigma)}$, $b^{(\sigma)}$,$d^{(\sigma)}$, and $h_{i,\sigma}$, is provided in Appendix~\ref{appendix:math}.
	
To capture the memory effects introduced by the capacitors \(C_p\) and \(C_n\), a transient simulation is performed by piecewise integrating the closed-form solutions across instantaneous modes. The state evolves as follows: \(\sigma\) is identified from the guard conditions (Table \ref{tab:modes_full}), \(V_p(t)\) is propagated via~\eqref{eq:Vp_solution}, and \(V_n(t)\) is recovered using~\eqref{eq:Vn_backsub}. When a transition between modes occurs, the constants \(C_1\) and \(C_2\) are updated to preserve the continuity of \(V_p(t)\) and \(\dot V_p(t)\). This ensures physically consistent trajectories across switching boundaries, reflecting the energy storage dynamics of the capacitors. A sufficiently small time step \(\Delta t\) is used to accurately track diode transitions and ripple effects. The above process is summarized in Algorithm~\ref{alg:dual_diode_sim}.

\begin{table}[t]
\centering
\caption{Conduction modes and guard conditions}
\label{tab:modes_full}
\scriptsize
\setlength{\tabcolsep}{1.5pt}
\renewcommand{\arraystretch}{1.2}

\rowcolors{2}{rowgray}{white} 
\begin{tabular}{c c c c c}
\toprule
Mode & $D_1,D_2$ & Guard Conditions
& $(R_{1,\sigma},R_{2,\sigma})$
& $(u_{1,\sigma}(t),u_{2,\sigma}(t))$ \\
\midrule
RR & Off, Off & $V_{D1}<V_{\rm on},\,V_{D2}<V_{\rm on}$ &
$(R_{\rm off},R_{\rm off})$ & $V_s(t),0$ \\
FR & On, Off & $V_{D1}\ge V_{\rm on},\,V_{D2}<V_{\rm on}$ &
$(R_{\rm on},R_{\rm off})$ & $V_s(t)-V_{\rm on},-V_{\rm on}$ \\
RF & Off, On & $V_{D1}<V_{\rm on},\,V_{D2}\ge V_{\rm on}$ &
$(R_{\rm off},R_{\rm on})$ & $V_s(t),-V_{\rm on}$ \\
FF & On, On & $V_{D1}\ge V_{\rm on},\,V_{D2}\ge V_{\rm on}$ &
$(R_{\rm on},R_{\rm on})$ & $V_s(t)-V_{\rm on}, 2V_{\rm on}$ \\
\bottomrule
\end{tabular}
\rowcolors{2}{}{} 
\end{table}
\section{Communication and Energy Harvesting}
\label{sec:comm_eh}
Let $V_L(t)$ denote the voltage across the load, equivalently expressed as the difference between the positive and negative rectified outputs, given by
\begin{equation}
    V_L(t) \triangleq V_p(t) - V_n(t).
\end{equation}
For information decoding, the U-SWIPT receiver performs sampling at the end of each symbol interval, $t = kT_s$, and define the noiseless state and its noisy observation as
\begin{equation}
    x_k = V_L(kT_s), \qquad
    y_k = x_k + w_k, \quad w_k \sim \mathcal{N}(0,\sigma^2),
\end{equation}
where $w_k$ accounts for additive Gaussian noise introduced by the sampling
and baseband processing.

\begin{figure}[t]
    \centering
    \includegraphics[width=\columnwidth]{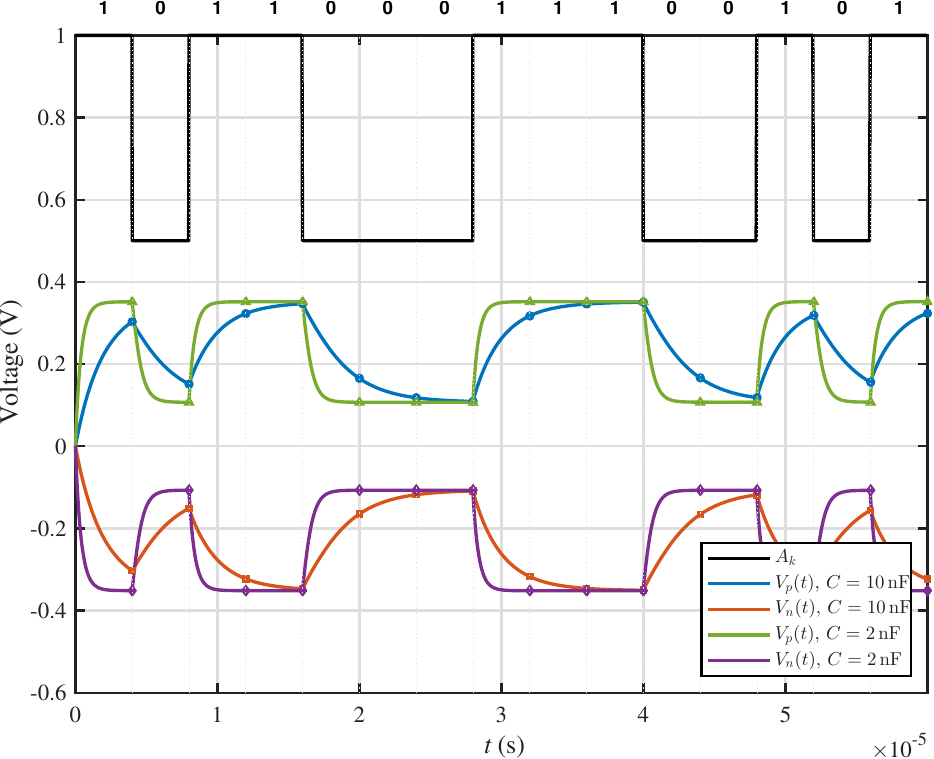}
    \caption{Transient response of the dual-diode U-SWIPT receiver for transmission
    of 15 bits using BASK with $T_s = 4~\mu\text{s}$. The digits above the trace show the
    transmitted bitstream. End-of-symbol samples ($t = kT_s^{-}$) are marked
    with circles/squares for the high-capacitance case ($C = 10~\text{nF}$)
    and triangles/diamonds for the low-capacitance case ($C = 2~\text{nF}$).
    Circuit parameters: $f = 800~\text{MHz}$, $R_s = 50~\Omega$,
    $R_L = 1~\text{k}\Omega$, $R_{\text{on}} = 5~\Omega$,
    $R_{\text{off}} = 10~\text{M}\Omega$, $V_{\text{on}} = 0.25~\text{V}$.}
    \label{fig:ddbaskdamwithbitstream}
\end{figure}

The memory behavior of the dual-diode receiver is governed by the effective
time constant
$T_0$, which characterizes the charging and
discharging of $C_p$ and $C_n$ through the source, load, and diode
resistances. When the capacitances are small, $T_0 \ll T_s$ and $V_L(t)$
approaches its steady-state value within each symbol, so $x_k$ depends only
weakly on past symbols and the receiver behaves nearly memoryless. As the capacitances increase, $T_0$ becomes comparable to or larger than $T_s$, preventing full settling between symbols. The residual charge stored on $C_p$ and $C_n$ then causes $x_k$ to depend strongly on the preceding bit sequence, yielding a nonlinear finite-memory channel induced purely by the rectifier dynamics. To illustrate how the capacitances $(C_p,C_n)$
control the transient behavior and the end-of-symbol samples $x_k$, 
Fig.~\ref{fig:ddbaskdamwithbitstream} shows the time-domain response of the
dual-diode receiver for two representative capacitance values at fixed
$T_s = 4~\mu\text{s}$.

\begin{algorithm}[t]
		\caption{Transient Simulation for the U-SWIPT Receiver}
		\label{alg:dual_diode_sim}
		\begin{algorithmic}[1]
			\State Initialize $t\!\gets\!0$, $\sigma\!\gets\!\mathrm{RR}$, $V_p,V_n\!\gets\!0$
			\For{$t=0:\Delta t:T_{\mathrm{sim}}$}
			\State Generate $V_s(t)$
			\State Evaluate diode voltages $V_{D1},V_{D2}$; determine new mode $\sigma$
			\State Propagate $V_p(t)$ via \eqref{eq:Vp_solution}; obtain $\dot V_p(t)$ and $V_n(t)$
			\If{mode switch}
			\State Update continuity constants $(C_1,C_2)$
			\EndIf
			\State Store $V_p(t),V_n(t)$
			\EndFor
		\end{algorithmic}
	\end{algorithm}

\subsection{State-Dependent Prediction Model}
\label{subsec:mod_sig}  

The symbol-to-symbol dependence induced by the rectifier memory can be compactly described through a deterministic
state-transition mapping obtained from the unified transient analysis of the receiver. When the $(k+1)$-th symbol is transmitted with amplitude $A_{k+1}$, the rectifier is excited with a sinusoidal waveform of amplitude $A_{k+1}$ over the interval $t \in [kT_s, (k+1)T_s)$. The resulting voltage at the end of the symbol, $V_L((k+1)T_s)$, is determined by the initial state $x_k$ and the applied symbol amplitude $A_{k+1}$. This deterministic relationship defines the state-transition function  

\begin{equation}
	x_{k+1} = f(x_k; A_{k+1}),
	\label{eq:f_transition}
\end{equation}
where $f(\cdot;\cdot)$ is obtained from the transient solution of the rectifier circuit over one symbol interval.

For the adopted binary BASK modulation with alphabet $\mathcal{A}=\{A_L, A_H\}$, the state evolution can be compactly described as a discrete-time switched system:
\begin{equation}
    x_{k+1} = 
    \begin{cases} 
        \mu_H(x_k) \triangleq f(x_k; A_H), & \text{if } A_{k+1} = A_H \\ 
        \mu_L(x_k) \triangleq f(x_k; A_L), & \text{if } A_{k+1} = A_L 
    \end{cases}
    \label{eq:switched_system}
\end{equation}
where $\mu_H(\cdot)$ and $\mu_L(\cdot)$ govern the deterministic state trajectories under high (`1') and low (`0') amplitude inputs, respectively. 
We precompute these functions offline over the invariant state domain $\mathcal{X} = [v_L, v_H]$, where $v_H$ ($v_L$) denotes the steady-state load voltage obtained under continuous excitation at amplitude $A_H$ ($A_L$). To facilitate efficient online detection, we discretize $\mathcal{X}$ into a uniform grid of $N$ points and tabulate the corresponding state transitions.

In the weak-memory regime, corresponding to small capacitances such that $T_0 \ll T_s$, the rectifier nearly settles within each symbol and the mappings become approximately constant,
$\mu_H(x) \approx v_H$ and $\mu_L(x) \approx v_L$ for all $x \in \mathcal{X}$. In contrast, for larger capacitances with $T_0 \gtrsim T_s$, incomplete charging and discharging render $\mu_H(x)$ and $\mu_L(x)$ nonlinear functions of the prior state $x$, encoding rectifier-induced memory.

\subsection{Detection Schemes}
Building upon the state-transition model, we now consider detection strategies for recovering transmitted symbols in the presence of rectifier-induced memory. In weak-memory regimes, a symbol-by-symbol maximum-likelihood (ML) detector with a fixed threshold is optimal, whereas in memory-dominated regimes the optimal receiver is maximum-likelihood sequence detection (MLSD), which in principle requires evaluating all $M^{K}$ possible symbol sequences~\cite{11015328}. This exponential complexity motivates the proposed circuit-aware scheme.

\subsubsection*{Circuit-Aware Adaptive Detection}
\label{subsubsec:circuit_aware}

To bridge the gap between the simple memoryless ML detector and the computationally prohibitive MLSD, we propose a \emph{Circuit-Aware Adaptive Detector (CAAD)} that explicitly exploits the deterministic memory of the rectifier for improved symbol recovery. By comparing the observation $y_k$ with these state-conditioned predictions, CAAD makes decisions that are inherently aware of the circuit's operating point, effectively turning memory into a source of decision context. The method uses precomputed mappings $\mu_H(x)$ and $\mu_L(x)$, obtained via offline transient simulation and stored for real-time access, thereby avoiding repeated numerical integration. At each symbol time, CAAD performs only two function evaluations and one comparison, yielding \emph{linear} complexity in block length, in stark contrast to the exponential complexity of MLSD.

Algorithm~\ref{alg:caad} summarizes the online operation. After an initialization phase based on a short pilot sequence that drives the rectifier to a known state $\hat{x}_0$, the detector proceeds symbol by symbol. At time $k$, CAAD first predicts the next end-of-symbol state for both hypotheses using the current state estimate, i.e., $\tilde{x}_{k+1|H} = \mu_H(\hat{x}_k)$ and
$\tilde{x}_{k+1|L} = \mu_L(\hat{x}_k)$. The received sample $y_{k+1}$ is then compared to these predictions, and the symbol whose prediction is closer in Euclidean distance is selected as $\hat{b}_{k+1}$. The state estimate is updated consistently with the decision, so that $\hat{x}_{k+1}$ always follows the chosen state-transition branch. In this way, CAAD incorporates the rectifier’s nonlinear memory into each decision using only two function evaluations and one comparison per symbol, resulting in linear complexity in the block length.

\begin{algorithm}[t]
\caption{Circuit-Aware Adaptive Detection (CAAD)}
\label{alg:caad}
\begin{algorithmic}[1]
\Require Precomputed mappings $\mu_H(\cdot),\mu_L(\cdot)$; pilot-based initial state $\hat{x}_0$; observations $\{y_k\}_{k=1}^{K}$
\Ensure Decoded bits $\{\hat{b}_k\}_{k=1}^{K}$
\For{$k = 0$ to $K-1$}
    \State $\tilde{x}_{k+1|H} \gets \mu_H(\hat{x}_k)$,\quad $\tilde{x}_{k+1|L} \gets \mu_L(\hat{x}_k)$
    \If{$|y_{k+1}-\tilde{x}_{k+1|H}| < |y_{k+1}-\tilde{x}_{k+1|L}|$}
        \State $\hat{b}_{k+1} \gets 1$,\quad $\hat{x}_{k+1} \gets \tilde{x}_{k+1|H}$
    \Else
        \State $\hat{b}_{k+1} \gets 0$,\quad $\hat{x}_{k+1} \gets \tilde{x}_{k+1|L}$
    \EndIf
\EndFor
\end{algorithmic}
\end{algorithm}

The offline computation of $\mu_H(x)$ and $\mu_L(x)$ requires running the transient solver (Algorithm~1) at a discrete set of initial conditions spanning $\mathcal{X}$. For a grid of $N$ points, this corresponds to $2N$ single-symbol simulations, performed once per circuit configuration. In contrast, MLSD must evaluate all $M^L$ possible sequences to decode each block, where $M$ is the modulation order and $L$ is the effective memory
length. For binary modulation with $L=10$, MLSD evaluates
$2^{10}=1024$ candidate trajectories per symbol, whereas CAAD evaluates exactly two.
Unlike the fixed-threshold ML detector, which uses $\rho = \tfrac{1}{2}(v_H + v_L)$ regardless of circuit history, CAAD
implicitly adapts its decision boundary based on the current state estimate $\hat{x}_k$. The effective threshold at symbol $k+1$ varies as the circuit state evolves through the transmission. This adaptation reflects the physical reality that the voltage swing induced by a `1' or `0' depends on the charge already present on the capacitors. As shown in 
Section~\ref{sec:results}, this memory-aware strategy allows CAAD to approach MLSD performance in memory-dominated regimes while maintaining the linear computational cost of symbol-by-symbol detection.

\subsection{Energy Harvesting}
\label{subsec:energy_harvesting}

We quantify the harvested energy via the power delivered to the load resistor $R_L$ at the sampling instants $t = kT_s$. For a given load voltage $v(t)$ used for joint ID/EH operation, the instantaneous harvested power is $P(t)=\frac{v^2(t)}{R_L}$, and over a sequence of $K$ symbols, the average harvested power is

\begin{equation}
    \bar{P}
    = \frac{1}{K} \sum_{k=1}^{K} P(kT_s)
    = \frac{1}{K R_L} \sum_{k=1}^{K} v^2(kT_s).
    \label{eq:avg_power}
\end{equation}

In the numerical results, \eqref{eq:avg_power} is evaluated for (i) $v(t)=V_p(t)$ and (ii) $v(t)=V_L(t)=V_p(t)-V_n(t)$, corresponding to the single-node and differential observation cases considered.

\section{Numerical Results}
\label{sec:results}

This section evaluates the performance of the proposed dual-diode U-SWIPT receiver in terms of information decoding and energy harvesting. Unless otherwise specified, simulations use the following parameters: carrier frequency $f_c = 800$~MHz, source resistance $R_s = 50~\Omega$, load resistance $R_L = 1$~k$\Omega$, diode turn-on voltage $V_{\text{on}} = 0.25$~V, forward resistance $R_{\text{on}} = 5~\Omega$, and reverse resistance $R_{\text{off}} = 10$~M$\Omega$. 

We consider biased binary amplitude shift keying (BASK) with symbol amplitudes  $A_L = 0.5~\text{V}$ and $A_H = 1.0~\text{V}$, ensuring $A_{\min} = 0.5~\text{V}$ and consistent modulation across all schemes. The received signal is corrupted by additive white Gaussian noise (AWGN) with variance $\sigma^2$, and the signal-to-noise ratio is expressed as $E_b/N_0$, where $\frac{E_b}{N_0} = \frac{P_{\mathrm{av}}}{2\sigma^2}$ and $P_{\mathrm{av}} = (A_L^2 + A_H^2)/2$ is the average transmit power per bit.

\begin{figure}[!t]
	\centering
	\includegraphics[scale=0.5]{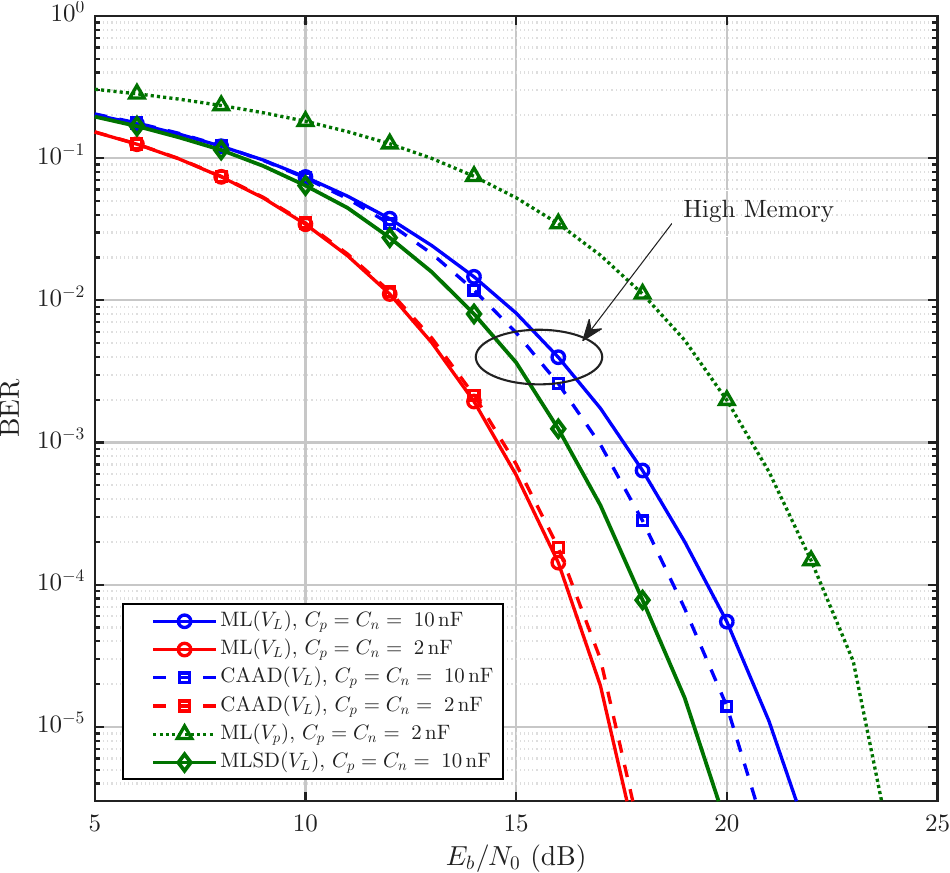}
	\caption{BER performance with respect to SNR for different schemes}
	\label{fig:BERplot}
\end{figure}

In Fig.~\ref{fig:BERplot}, we plot the BER performance versus $E_b/N_0$ for the dual-diode receiver under four detection schemes: symbol-by-symbol ML using $V_p(t)$, symbol-by-symbol ML using $V_L(t)$, the proposed CAAD operating on $V_L$, and MLSD as a benchmark. The MLSD curve is obtained by applying maximum-likelihood sequence detection with an effective memory
length $L = 10$, i.e., by evaluating all $2^{L}$ candidate binary sequences
generated by the transient model for each block. For $C_p = C_n = 2$ nF, all schemes exhibit practically identical performance, which confirms that the circuit operates in a weak-memory regime where the steady-state model is sufficient and a fixed-threshold ML detector is adequate. For $C_p = C_n = 10$ nF, the increased memory causes the sampled outputs to cluster in state-dependent regions; in this case both ML detectors suffer from an error floor, while CAAD closely follows the MLSD curve over the entire SNR range. These results verify that CAAD effectively exploits the deterministic state transitions $\mu_H(x)$ and $\mu_L(x)$ to account for rectifier memory, achieving near-optimal sequence detection behavior without the exponential complexity associated with MLSD.

\begin{table}[!t]
\caption{Average Harvested Power}
\label{tab:avg_power_bar}
\centering
\begin{tabular}{|c|c|c|}
\hline
\textbf{Capacitance} & $\overline{P}(V_L)$ ($\mu$W) & $\overline{P}(V_p)$ ($\mu$W) \\
\hline
$C_p = C_n = 2.0$~nF & 247.9 & 61.9 \\
\hline
$C_p = C_n = 10.0$~nF & 248.1 & 62.0 \\
\hline
\end{tabular}
\end{table}

Finally, Table~\ref{tab:avg_power_bar} presents the average harvested power for both capacitance values and observation nodes, computed using \eqref{eq:avg_power}. Larger capacitances yield a slightly higher average harvested power, since the increased storage prevents the load voltage from decaying toward the lower steady-state between symbols. However, this dependence is marginal for the considered values of $(C_p,C_n)$, indicating that capacitance primarily affects transient memory and ID performance rather than long-term energy throughput. In contrast, the harvesting topology plays a decisive role. Using the differential node $V_L(t)$ delivers significantly higher harvested power compared to the single-ended node $V_p(t)$ (e.g., $247.9~\mu$W vs. $61.9~\mu$W). This gain arises from the dual-diode configuration exploiting both waveform polarities, effectively increasing the load voltage swing and, hence, the harvested power.

\section{Conclusions}
This paper presented a unified transient framework for a dual-diode unified SWIPT receiver that jointly accounts for diode nonlinearity and capacitor-induced memory. The proposed analytical model provides an accurate and tractable characterization of the symbol-to-symbol behavior of the rectifier and serves as a common basis for evaluating both information detection and energy harvesting performance. With this framework, we introduced a circuit-aware adaptive detection scheme that exploits precomputed state-transition mappings to perform decision-directed symbol detection with linear complexity. Numerical results demonstrated that the proposed detector achieves performance close to maximum-likelihood sequence detection in memory-dominated regimes, while substantially outperforming conventional memoryless detection schemes, and that the dual-diode architecture offers an improved joint rate–reliability–energy tradeoff compared to conventional unified receiver designs.

\appendices
\section{Proofs of Closed-Form Expressions in Eqs.~\eqref{eq:Vp_solution} $\&$\eqref{eq:Vn_backsub}}
\label{appendix:math}
Starting from the coupled ODEs in \eqref{eq:coupled1}–\eqref{eq:coupled2}, we define the auxiliary coefficients:
\begin{align*}
A_\sigma &= (R_s + R_{1,\sigma})C_p, &
B_\sigma &= R_s C_n, \\
C_\sigma &= 1 + \tfrac{R_{1,\sigma}}{R_L}, &
D_\sigma &= -\tfrac{R_{1,\sigma}}{R_L},
\end{align*}

\begin{align*}
E_\sigma &= R_{1,\sigma}C_p, &
F_\sigma &= -R_{2,\sigma}C_n, \\
G_\sigma &= \tfrac{R_{1,\sigma} + R_L + R_{2,\sigma}}{R_L}, &
H_\sigma &= -G_\sigma.
\end{align*}

and the determinant $\Delta_\sigma = B_\sigma H_\sigma - D_\sigma F_\sigma$.
Eliminating $V_n(t)$ yields the scalar second-order ODE
\begin{equation}
	K_1^{(\sigma)} \ddot V_p(t) + K_2^{(\sigma)} \dot V_p(t) + K_3^{(\sigma)} V_p(t) = g^{(\sigma)}(t),
	\label{eq:app_scalarODE}
\end{equation}
with coefficients
\begin{equation}
	\begin{aligned}
		K_1^{(\sigma)} &= \frac{B_\sigma (A_\sigma F_\sigma - B_\sigma E_\sigma)}{\Delta_\sigma}, \\
		K_2^{(\sigma)} &= \frac{B_\sigma (A_\sigma H_\sigma - B_\sigma G_\sigma + C_\sigma F_\sigma - D_\sigma E_\sigma)}{\Delta_\sigma}, \\
		K_3^{(\sigma)} &= \frac{B_\sigma (C_\sigma H_\sigma - D_\sigma G_\sigma)}{\Delta_\sigma}.
	\end{aligned}
	\label{eq:app_K123}
\end{equation}

For a single-tone input $V_s(t) = \bar{V}_s \sin(2\pi f t)$, the general solution is
\begin{align}
	V_p(t) = &\, C_1 e^{r_1^{(\sigma)}(t - t_0)} + C_2 e^{r_2^{(\sigma)}(t - t_0)} \notag \\
	&+ \bar{V}_s \bigl[ a^{(\sigma)} \cos(2\pi f t) + b^{(\sigma)} \sin(2\pi f t) \bigr] + d^{(\sigma)}, \vspace{-2mm}
	\label{eq:app_Vp_sol}
\end{align}
where $r_{1,2}^{(\sigma)}$ are roots of $K_1^{(\sigma)} \lambda^2 + K_2^{(\sigma)} \lambda + K_3^{(\sigma)} = 0$, and the harmonic gains are
\begin{equation}
	\begin{aligned}
		a^{(\sigma)} &= \frac{-K_1^{(\sigma)} \omega^2 + K_3^{(\sigma)}}{\bigl(-K_1^{(\sigma)} \omega^2 + K_3^{(\sigma)}\bigr)^2 + \bigl(K_2^{(\sigma)} \omega\bigr)^2}, \\
		b^{(\sigma)} &= \frac{K_2^{(\sigma)} \omega}{\bigl(-K_1^{(\sigma)} \omega^2 + K_3^{(\sigma)}\bigr)^2 + \bigl(K_2^{(\sigma)} \omega\bigr)^2}.
	\end{aligned}
	\label{eq:app_ab}
\end{equation}

The mode-dependent DC offsets in \eqref{eq:app_Vp_sol} are:
\begin{equation}
	\begin{aligned}
		d^{(\mathrm{RR})} &= 0, \\
		d^{(\mathrm{FR})} &= \frac{1}{K_3^{(\mathrm{FR})}} \left( \frac{D_{\mathrm{FR}}(B_{\mathrm{FR}} - F_{\mathrm{FR}}) - B_{\mathrm{FR}} H_{\mathrm{FR}}}{\Delta_{\mathrm{FR}}} \right) V_{\mathrm{on}}, \\
		d^{(\mathrm{RF})} &= \frac{1}{K_3^{(\mathrm{RF})}} \left( \frac{D_{\mathrm{RF}} \left(1 - \frac{F_{\mathrm{RF}}}{B_{\mathrm{RF}}} \right)}{H_{\mathrm{RF}} - \frac{F_{\mathrm{RF}} D_{\mathrm{RF}}}{B_{\mathrm{RF}}}} - 1 \right) V_{\mathrm{on}}, \\
		d^{(\mathrm{FF})} &= \frac{1}{K_3^{(\mathrm{FF})}} \left( \frac{D_{\mathrm{FF}}(B_{\mathrm{FF}} - F_{\mathrm{FF}})}{\Delta_{\mathrm{FF}}} - 1 \right) 2V_{\mathrm{on}}.
	\end{aligned}
	\label{eq:app_d_offsets}
\end{equation}
The harmonic gains $a^{(\sigma)}$ and $b^{(\sigma)}$ in \eqref{eq:app_ab} are obtained by solving the scalar ODE \eqref{eq:app_scalarODE} under sinusoidal excitation; their closed-form expressions follow from standard undetermined-coefficients analysis.

The back-substitution formula for $V_n(t)$ is
\begin{equation}
	V_n(t) = h_{1,\sigma} V_p(t) + h_{2,\sigma} \dot V_p(t) + h_{3,\sigma} V_s(t) + h_{4,\sigma},
	\label{eq:app_Vn}
\end{equation}
with coefficients given by
\begin{equation}
	\begin{aligned}
		h_{1,\sigma} &= -\frac{G_\sigma - \frac{F_\sigma C_\sigma}{B_\sigma}}{H_\sigma - \frac{F_\sigma D_\sigma}{B_\sigma}}, &
		h_{2,\sigma} &= -\frac{E_\sigma - \frac{F_\sigma A_\sigma}{B_\sigma}}{H_\sigma - \frac{F_\sigma D_\sigma}{B_\sigma}}, \\
		h_{3,\sigma} &= -\frac{F_\sigma}{B_\sigma \left( H_\sigma - \frac{F_\sigma D_\sigma}{B_\sigma} \right)}, &
		h_{4,\sigma} &= -\frac{u_{2,\sigma}^{(0)} - \frac{F_\sigma}{B_\sigma} u_{1,\sigma}^{(0)}}{H_\sigma - \frac{F_\sigma D_\sigma}{B_\sigma}},
	\end{aligned}
	\label{eq:app_h_direct}
\end{equation}
where $(u_{1,\sigma}^{(0)}, u_{2,\sigma}^{(0)})$ are the constant (threshold) terms from Table~I.

At each mode transition, continuity of voltage and current across the filter
capacitors requires $V_p(t)$ and $\dot V_p(t)$ to remain continuous at the
switching instant $t_0$.  Let $V_{p0}\!\triangleq V_p(t_0)$ and
$\dot V_{p0}\!\triangleq \dot V_p(t_0^+)$ denote the post–switch values.
The particular (steady–state) component of~\eqref{eq:app_Vp_sol} is
\begin{align}
p_\sigma(t) &= \bar{V}_s
\big[a^{(\sigma)}\cos(2\pi f t) + b^{(\sigma)}\sin(2\pi f t)\big] + d^{(\sigma)}, \nonumber\\
\dot p_\sigma(t) &= \bar{V}_s
\big[-a^{(\sigma)}(2\pi f)\sin(2\pi f t)
+ b^{(\sigma)}(2\pi f)\cos(2\pi f t)\big], \nonumber
\end{align}
so that $p_\sigma(t_0)$ and $\dot p_\sigma(t_0)$
represent the steady–state voltage and its slope at the mode–entry time.

Matching $V_p$ and $\dot V_p$ at $t_0$ yields the transient constants
\begin{equation}
\begin{aligned}
C_1 &= \frac{\dot V_{p0}-\dot p_\sigma(t_0)
- r_2^{(\sigma)}\!\big(V_{p0}-p_\sigma(t_0)\big)}
{r_1^{(\sigma)}-r_2^{(\sigma)}},\\[0.3em]
C_2 &= \frac{r_1^{(\sigma)}\!\big(V_{p0}-p_\sigma(t_0)\big)
- \big(\dot V_{p0}-\dot p_\sigma(t_0)\big)}
{r_1^{(\sigma)}-r_2^{(\sigma)}}.
\end{aligned}
\label{eq:app_C1C2}
\end{equation}
These expressions ensure smooth evolution of $V_p(t)$ and $\dot V_p(t)$
across consecutive conduction modes.

\bibliographystyle{IEEEtran}
\bibliography{refs}
\end{document}